\documentclass[preprint2,twoside]{hwo}

\usepackage{lipsum}

\input{hwo.h}

\begin{document}

\title{\textbf{\LARGE How Do Ionizing Photons Escape from Star-Forming Galaxies?}}
\author {\textbf{\large Cody Carr,$^{1,2}$ Renyue Cen,$^{1,2}$ Sophia Flury,$^{3,4}$ M. S. Oey,$^5$ Stephan McCandliss,$^6$ \\ Allison Strom,$^{7,8}$}}
\affil{$^1$\small\it Center for Cosmology and Computational Astrophysics, Institute for Advanced Study in Physics \\ Zhejiang University, Hangzhou 310058,  China, \email{codycarr24@gmail.com,renyuecen@zju.edu.cn}}
\affil{$^2$\small\it Institute of Astronomy, School of Physics, Zhejiang University, Hangzhou 310058,  China}

\affil{$^3$\small\it Department of Astronomy, University of Massachusetts, Amherst,MA 01003, USA}
\affil{$^4$\small\it Institute for Astronomy, University of Edinburgh, Royal Observatory, Edinburgh, EH9 3HJ, UK}
\affil{$^5$\small\it Astronomy Department, University of Michigan, Ann Arbor, MI 48109, USA}
\affil{$^6$\small\it Johns Hopkins University, Department of Physics \& Astronomy, Center for Astrophysical Sciences, 3400 North Charles Street, Baltimore, MD, 21218, USA}
\affil{$^7$\small\it Center for Interdisciplinary Exploration and Research in Astrophysics (CIERA), Northwestern University, 1800 Sherman Ave., Evanston, IL 60201, USA}
\affil{$^8$\small\it Department of Physics and Astronomy, Northwestern University, 2145 Sheridan Road, Evanston, IL 60208, USA}



\author{\footnotesize{\bf Endorsed by:}
Logan Jones (Space Telescope Science Institute), Jean-Claude Bouret (Laboratoire d'Astrophysique de Marseille), Nikole Nielsen (University of Oklahoma), Mason Huberty (University of Minnesota), Marc Rafelski (Space Telescope Science Institute), Luca Fossati (Space Research Institute, Austrian Academy of Sciences), Eunjeong Lee (EisKosmos (CROASAEN), Inc.), Annalisa Citro (University of Minnesota), Kevin France (University of Colorado Boulder), Brad Koplitz (Arizona State University), Michelle Berg (Texas Christian University), Ariane Lançon (Observatoire astronomique de Strasbourg - France), Sylvain Veilleux (University of Maryland, College Park), Lukas Furtak (Ben-Gurion University of the Negev), and Laura Pentericci (INAF-Astronomical Observatory of Rome)
}

\begin{abstract}
  
The Epoch of Reionization marks the last major phase transition in the early Universe, during which the majority of neutral hydrogen once filling the intergalactic medium was ionized by the first galaxies. The James Webb Space Telescope is now identifying promising galaxy candidates capable of producing sufficient ionizing photons to drive this transformation. However, the fraction of these photons that escape into intergalactic space—the escape fraction—remains highly uncertain. Stellar feedback is thought to play a critical role in carving low-density channels that allow ionizing radiation to escape, but the dominant mechanisms, their operation, and their connection to observable signatures are not well understood. Local analogs of high-redshift galaxies offer a powerful alternative for studying these processes, since ionizing radiation is unobservable at high redshift due to intergalactic absorption. However, current UV space-based instrumentation lacks the spatial resolution and sensitivity required to fully address this problem. The core challenge lies in the multiscale nature of LyC escape: ionizing photons are generated on scales of 1–100 pc in super star clusters but must traverse the circumgalactic medium which can extend beyond 100 kpc. A UV integral field unit (IFU) spectrograph capable of resolving galaxies across these scales is necessary—and uniquely achievable with the proposed Habitable Worlds Observatory. In this article, we outline the scientific motivation, observables, and observational capabilities needed to make progress on these fundamental questions.
  \\
  \\
\end{abstract}

\vspace{2cm}

\section{Introduction}


\begin{figure*}[ht]
\begin{center}
\includegraphics[width=\textwidth]{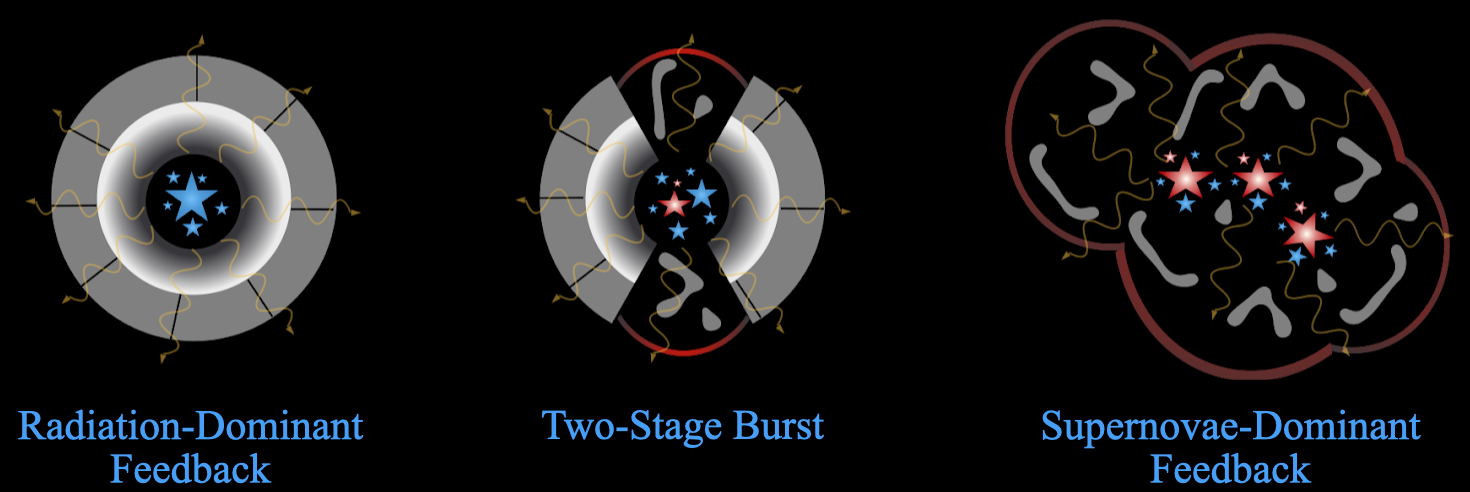}
\caption{\small How does feedback shape the neutral gas and dust content of galaxies?  \textbf{Left}: In young stellar clusters, radiation pressure and ionizing fronts evacuate low-density channels, which form naturally from turbulence and cloud fragmentation in the ISM \citep{Kakiichi2021}. \textbf{Middle}: During the onset of supernovae, mechanical feedback helps clear channels which are further evacuated by intense radiation fronts in a “two-stage burst” scenario \citep{Martin2024,Flury2022a,Flury2025}.  \textbf{Right}: In evolved stellar populations, supernova blastwaves may lift neutral gas and dust, clearing pathways for LyC escape \citep{Kimm2014,Cen2020}.
\label{fig:feedback}
}
\end{center}
\end{figure*}

Approximately one billion years after the Big Bang, the Universe experienced its last major phase transition, during which the majority of neutral hydrogen once filling the intergalactic medium (IGM) became ionized. This era, known as the Epoch of Reionization (EoR), remains one of the last major frontiers in cosmology \citep{Robertson2022}. Key questions, such as the exact duration of the EoR and the identity of the astrophysical objects responsible for reionizing the Universe, remain unanswered. 

Early results from the James Webb Space Telescope (JWST) suggest that star-forming (SF) galaxies produce enough ionizing, or Lyman continuum (LyC; $< 912 $ \AA), photons to reionize the Universe \citep{Atek2024,Lin2024,Munoz2024,Pahl2025}. However, it remains unclear how many of these photons actually escape from the first galaxies to ionize the IGM. Neutral gas and dust within the interstellar (ISM) and circumgalactic medium (CGM) efficiently absorb LyC photons, making escape unlikely without some form of intervention. Feedback processes—such as radiation pressure, stellar winds, and supernovae—are thought to create low-density channels that allow LyC photons to leak out. Yet, the dominant mechanism, how these processes operate, and how they connect to observable signatures are still open questions. Figure 1 illustrates three possible feedback-driven escape scenarios.

Deciphering the physical mechanisms that govern LyC escape is inherently a multiscale problem—ranging from the production of LyC photons in super star clusters (SSCs; $\sim$1–100 pc) to their eventual escape through the CGM ( $\sim$1–100 kpc). Currently, the astronomical community lacks the space-based instrumentation necessary to resolve ultraviolet emission on SSC scales. This limitation hinders our understanding of how stellar feedback shapes the distribution of neutral gas and dust in the CGM and ultimately limits our ability to determine how SF galaxies contributed to reionization

These challenges have been broadly recognized within the astronomical community and are directly aligned with the key scientific priorities outlined in the 2020 Astrophysics Decadal Survey \citep{nasem2023}, including:

\begin{itemize}
  \item \textbf{D-Q1:} How did the intergalactic medium and the first sources of radiation evolve from cosmic dawn through the epoch of reionization?
  \item \textbf{D-Q2:} How do gas, metals, and dust flow into, through, and out of galaxies?
\end{itemize}

\section{Science Objectives}

\begin{figure*}[ht]
\begin{center}
\includegraphics[width=\textwidth]{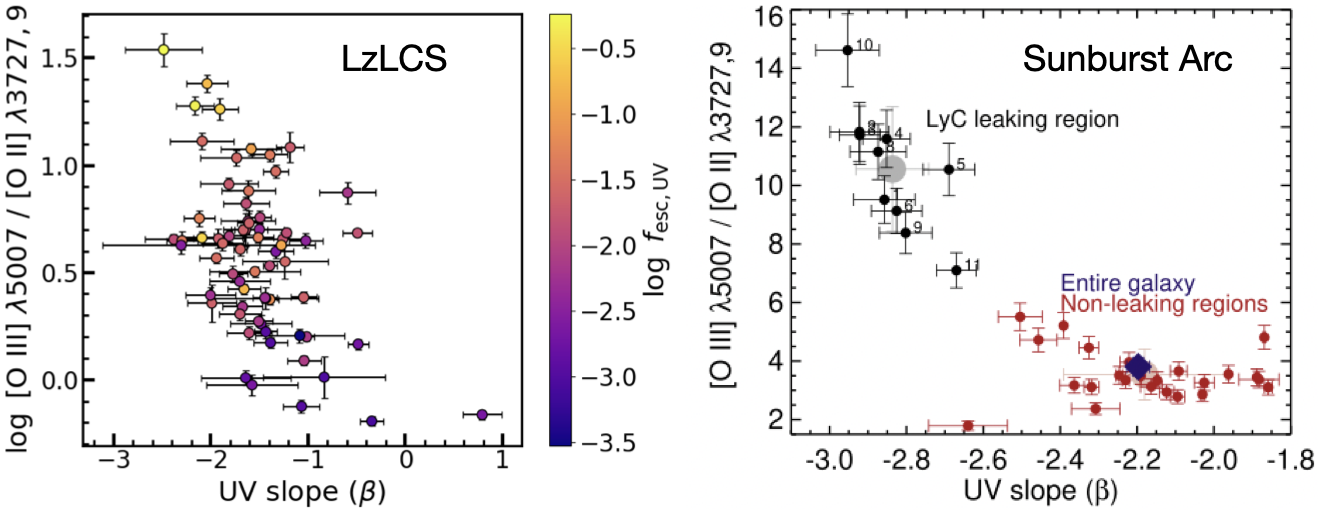}
\caption{\small LyC escape according to the [O III] 5007\AA\ / [O II] 3726,9\AA\ ratio and the slope of the FUV continuum slope ($\beta$).  The left panel shows values derived from HST COS integrated spectra for entire galaxies selected from LzLCS.  Data were taken from \cite{Flury2022a,Flury2022_erratum}.  The right panel shows values drawn from individual regions of the Sunburst Arc galaxy, figure taken from \cite{Kim2023}.  The black points correspond to regions with LyC emission and the red points without.  The blue point represents the average value measured over the whole galaxy.  To understand the relationship between $f_{\mathrm{esc}}^{\mathrm{LyC}}$ and observables, we must resolve galaxies on the cluster scale.
\label{fig:diagnostics}
}
\end{center}
\end{figure*}

\begin{figure*}[ht]
\begin{center}
\includegraphics[width=\textwidth]{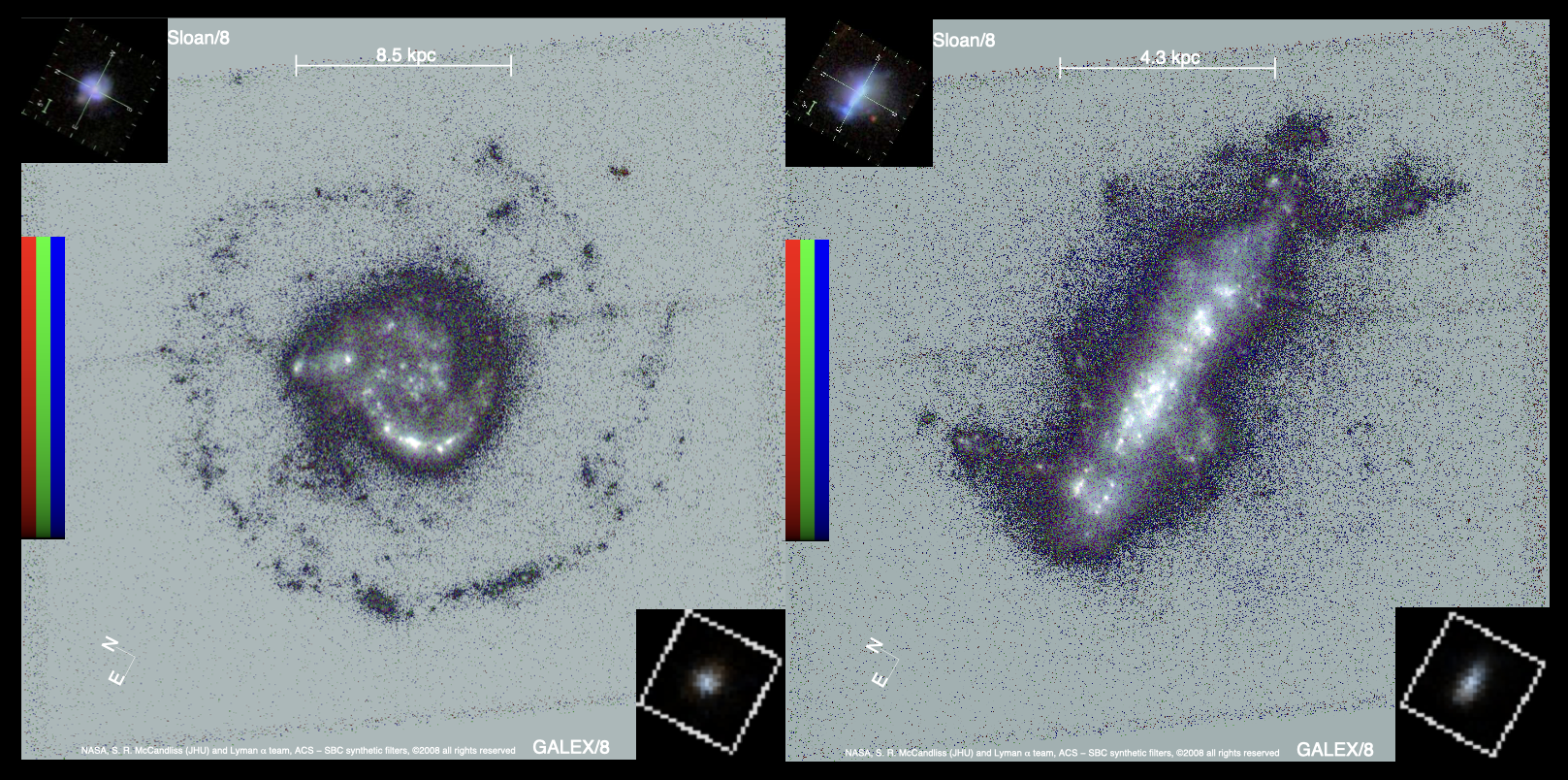}
\caption{\small ACS/SBC synthetic filter images of two nearby star-forming galaxies \citep{McCandliss2009}.  Blue subplots show regions of diffuse Ly$\alpha$ emission.  We count roughly 20 (left panel) and 30 (right panel) distinct star clusters in each galaxy.  Spatially resolving such features will be critical for understanding how LyC photons escape from galaxies and linking them to observational signatures.  
\label{fig:clusters}
}
\end{center}
\end{figure*}


\begin{itemize}
    \item How is LyC escape linked to various galaxy properties and high-\textit{z} observables?  
    \item What is the dominant source of feedback governing LyC escape across different galaxy populations? 
    \item How does feedback (mechanical vs. radiation) shape the distribution of neutral gas and dust in galaxies to allow LyC escape?  
    \item How does the LyC escape fraction evolve relative to the star formation episode or baryon cycle?  
    \item What is the appropriate timeline for LyC escape? 
    \item What are the roles of different stellar populations?
\end{itemize}

The primary challenge to measuring a galaxy’s contribution to reionization lies in determining how much of its intrinsic LyC production actually goes on to ionize the IGM.  This quantity is referred to as the LyC escape fraction ($f_{\mathrm{esc}}^{\mathrm{LyC}}$).  Absorption in the neutral IGM prevents direct measurements of $f_{\mathrm{esc}}^{\mathrm{LyC}}$ at high redshifts \citep{Inoue2014}.  Consequently, astronomers must rely on local analogs of high redshift galaxies and simulations to develop diagnostics of $f_{\mathrm{esc}}^{\mathrm{LyC}}$ to infer the contributions of individual galaxies to reionization \citep{Flury2022b,Choustikov2024,Jaskot2024a,Jaskot2024b}.  For instance, high [O III] 5700\AA/ [O II] 3726,9\AA\ ratios (O32; \citealt{Jaskot2013,Nakajima2014,Izotov2017,Izotov2018b,Izotov2020}) and the non-ionizing UV continuum slope ($\beta_{\mathrm{UV}}$; \citealt{Zackrisson2013,Zackrisson2017,Chisholm2022}) have all been linked to $f_{\mathrm{esc}}^{\mathrm{LyC}}$ in local galaxies.  These quantities contain information about the ionizing intensity and dust content, respectively.  The left panel of Figure 2 shows how $f_{\mathrm{esc}}^{\mathrm{LyC}}$ varies with O32 and UV in local galaxies drawn from the Low-z Lyman Continuum Survey (LzLCS; \citealt{Flury2022a}) - the largest study of local ($0.2 < z < 0.4$) LyC emitters to-date.  Note the large variation towards the middle of the diagram.  The consensus of this and similar studies is that no single perfect indicator exists, reflecting the complex physics governing the escape of LyC photons from galaxies and their immediate surroundings \citep{Flury2022b,Jaskot2024b}.

Most studies investigating LyC escape diagnostics rely on spatially integrated spectra, such as those obtained with the Cosmic Origins Spectrograph (COS) aboard the Hubble Space Telescope (HST), and therefore do not capture spatial variations within galaxies. In reality, LyC production is highly localized, occurring primarily within SSCs composed of massive O- and B-type stars—the key sources of ionizing radiation. The Sunburst Arc, a strongly lensed LyC-emitting galaxy, offers a rare opportunity to study LyC escape at sub-galactic scales, approaching the level of individual clusters. The right panel of Figure 2 illustrates how LyC emission in the Sunburst Arc varies significantly across its face along with the $\beta_{UV}$ and the O32 ratio. The diagnostics differ substantially between LyC-emitting and non-emitting regions, with O32 generally increasing in regions showing LyC emission \citep{Kim2023}.  However, this physical connection between O32 and $f_{\mathrm{esc}}^{\mathrm{LyC}}$ is lost when averaging over the full galaxy, as is done in integrated spectroscopy.  Therefore, to meaningfully link LyC escape with observable diagnostics—and to understand the physical mechanisms behind their connection—analyses must be performed at the cluster scale.

For LyC photons to escape from galaxies into the IGM, SSCs must clear their immediate surroundings of neutral hydrogen and dust. This clearing is driven by various feedback processes, including radiation pressure, stellar winds, and supernova explosions. Ram pressure from supernova-driven blast waves, for example, can launch massive, multiphase outflows that lift LyC-blocking clouds from the ISM \citep{Kimm2014,Cen2020}, while intense ionizing radiation from young stars may create escape pathways by ionizing low-column-density channels which naturally occur in turbulent media \citep{Kakiichi2021,Menon2024}. The dominant feedback mechanism operating in a given region can be inferred from the age of the stellar populations—for instance, radiation feedback typically dominates in clusters younger than 6 Myr \citep{Flury2025,Carr2025LyC}.  Current observational evidence suggests that LyC escape is highest during the earliest phases of star formation, when young, massive stars emit strong ionizing radiation prior to the onset of supernovae \citep{Hayes2023b,Bait2024,Flury2025,Carr2025LyC}.  In contrast, simulations often favor later evolutionary stages, when supernova-driven winds have had sufficient time to evacuate dense, LyC-blocking gas \citep{Kimm2014,Cen2020,Rosdahl2022,Choustikov2024}.  Disentangling these mechanisms will require spectral imaging of the winds, down to the scale of the SSCs.   
 
The kinematic and geometric (density + distribution) properties of the different outflow phases needed to understand how winds shape their environments is reflected in the broadening of absorption \citep{Prochaska2011,Scarlata2015,Zhu2015,Chisholm2016b,Carr2018} and emission lines \citep{Flury2023,Yuan2023,Amorin2024}.  While empirical measures such as line widths can be recovered from low-resolution spectra (R $<$ 5000), more nuanced parameters—such as density and velocity gradients—require high spectral resolution (R $\sim 15000$, or $\sim$ 20 $\mathrm{km\,s^{-1}}$) for effective radiative transfer modeling \citep{Carr2023,Carr2025MOR}.  However, spatially resolved spectra such as that obtainable with integral field unit (IFU) spectroscopy provides the most promising data set for constraining radial profiles \citep{Burchett2021,Erb2023,Carr2025MOR}.  Furthermore, mapping the winds on a cluster-by-cluster basis, as possible with both a multi-object spectrograph (MOS) or IFU would allow one to pair the feedback mechanism to the locations of LyC escape.  Figure~\ref{fig:clusters} shows synthetic filters of the star-forming knots in two local star-forming galaxies, suggesting a target sample of a few dozen SSCs per galaxy \citep{McCandliss2009}.  Together, these advances would enable a definitive identification of the physical mechanisms driving LyC escape in the local Universe and establish robust connections to high-redshift diagnostics, thereby clarifying the contribution of star-forming galaxies to reionization.

Understanding how much LyC photons (Q) different stellar populations contribute to the LyC budget is also a crucial component to understanding how LyC photons escape from SF galaxies.  This is typically done through modeling of the spectral energy distributions (SEDs; \citealt{Steidel2016,Chisholm2019,Chisholm2022,Flury2025}).  This can be done at lower resolution R$\sim 3000-5000$. Other relevant stellar properties include the metallicity and star formation history. 


\section{Physical Parameters}

The following list of parameters provides useful information as to how star formation shapes the structure of the ISM and CGM.

\begin{itemize}
  \item the Lyman continuum escape fraction $f_{\mathrm{esc}}$
  \item the properties of cool and warm-hot galactic outflows and inflows
  \begin{itemize}
    \item mass, momentum, energy outflow and inflow rates
    \item terminal velocity
    \item column density
    \item spatial distribution and extent
    \item density and velocity fields
    \item turbulent velocity dispersion
  \end{itemize}
  \item the properties of stellar populations
    \begin{itemize}
    \item metallicity
    \item age
    \item star formation history
    \item total ionizing production ($Q$)
    \end{itemize}
 
  \item $\mathrm{O\ [III]}\ 5007\ \mathrm{\AA}\ /\ \mathrm{O\ [II]}\ 3727, 29 \ \mathrm{\AA}$ ratios 
  \item UV brightness profile
  \item non-ionizing FUV continuum slope, $\beta$
  \item galaxy properties
  \begin{itemize}
    \item stellar mass
    \item metallicity (gas and stars)
    \item SFR
    \item morphology
  \end{itemize}
\end{itemize}

\section{Description of Observations}

To understand how LyC radiation escapes from galaxies, we must simultaneously study the LyC, H I, and UV metal line emission from individual SSCs.  This amounts to a few dozen SSCs per galaxy (see Figure~\ref{fig:clusters}).  By studying a range of UV lines, spanning different ionization states, we will be able to study the multiphase nature of galactic winds as well as the structure of the ISM.  All further claims assume a standard flat cosmology:  $H_0 = 70$ km s${}^{-1}$ Mpc${}^{-1}$, $\Omega_m=0.3$, $\Omega_{\lambda}=0.7$.

\subsubsection{Lyman Continuum}

The LyC escape fraction can be measured directly from emission at wavelengths below the 912 \AA\ Lyman limit.  For instance, an empirical proxy is commonly used, where the flux near 900 \AA\ is compared to the continuum flux around 1100~\AA\ \citep{Wang2019,Flury2022a}.  Current constraints on $f_{\mathrm{esc}}^{\mathrm{LyC}}$ have been measured using low-resolution spectroscopy ($\rm R \sim 1050$ at 1100 \AA) with the COS G140L grating, which corresponds to an observed wavelength of 900 \AA\ at $z = 0.22$, typically at a signal-to-noise ratio of $\rm S/N = 5$ \citep{Flury2022a}.  The COS throughput sensitivity drops by roughly two-orders of magnitude below 1100 \AA, imposing a lower limit at this redshift \citep{Green2012}.  Because we desire to spatially resolve LyC emission on the SSC scale (1-100 pc), the closer we can observe an object the better when observing at low redshifts.  The diffraction limit of a 6(8)-meter telescope is 4.6(3.5) mas at 1100 \AA.  At three times the diffraction limit, we can resolve roughly a 49(37) pc object at redshift $z = 0.22$.  If we look closer at  $z = 0.1$, this corresponds to distances of 26(19) pc.  However, contamination from Milky Way absorption will become more and more problematic at lower redshifts.  For instance, the 900 \AA\ target will be redshifted to 990 \AA\ at $z = 0.1$.  While this wavelength is safe from absorption by Milky Way hydrogen below the Lyman edge (912 \AA), we still have to worry about the abundance of absorption lines appearing just beyond the edge (e.g., L$\beta$, L$\gamma$, etc.) making clean observations of the LyC at $z = 0.1$ difficult. 
 
While most studies focus on stellar contributions to the LyC, in reality, there may be contributions coming from nebular emission as well - for instance, free electrons recombining to the first excited state of hydrogen will create a flux excess just below the Lyman limit at 912 \AA, often called the “Lyman bump” \citep{Inoue2010}.  Since the nebular contribution is anticipated to be highly wavelength dependent, it may influence the perceived value for $f_{\mathrm{esc}}^{\mathrm{LyC}}$.  To ensure separation from the nebular contribution, a broader wavelength range of 700–912 \AA\ would be ideal, as the 700 \AA\ mark is likely far enough away from the nebular contributions from higher order H I transitions \citep{Simmonds2024}.  If we extend to even bluer wavelengths, down to 400 \AA, we would also be able to measure the He Lyman bump below its Lyman limit at 504 \AA.  This must be done at much higher redshifts, however.  For example, 400 \AA\ would be observed at 1100 \AA\ at $z = 1.75$.  Increasing sensitivity below 1100 \AA\ would help, but again Milky Way contamination plays a factor.  A resolution of $\rm R = 1000-5000$ and $\rm S/N = 5$ over the rest wavelength range 400 - 912 \AA\ would be ideal for resolving the slope of the anticipated nebular contributions.  

Flux sensitivity is also an important property to consider when measuring $f_{\mathrm{esc}}^{\mathrm{LyC}}$.  As a lower limit, the observed flux (accounting for IGM absorption) at 900 \AA\ is projected to be about $10^{-19}$ erg/s/cm$^2$/\AA\ or about 30 absolute magnitudes for $f_{\rm esc}^{\rm LyC} = 0.3\%$ in typical SF galaxies.  These numbers drop by about one dex out to $z = 1$ and by two dex at $z = 3$ \citep{McCandliss2017}.      

The following lines will be useful in the study of galactic winds.

\subsubsection{Lyman Alpha}

The 1216 \AA\ Lyman alpha (Ly$\alpha$) line is the best known proxy of LyC emission.  A double peaked emission feature is often observed in SF galaxies with the peak separation showing a negative correlation with $f_{\rm esc}^{\rm LyC}$ \citep{Flury2022b,Izotov2016a,Izotov2016b,Izotov2018b,Izotov2021,Izotov2022}.  Ly$\alpha$ acts as a direct and sensitive probe of the neutral gas and dust distribution in galaxies and is a prime tracer of the bulk of mass carried in galactic outflows \citep{Verhamme2015,Chung2019}.  The Ly$\alpha$ absorption feature has been linked to the high velocity profile of various metal lines \citep{Erb2023}.  Rather surprisingly, Ly$\alpha$ emission is frequently observed at high redshifts \citep{Saxena2024}, likely following the escape through ionized bubbles \citep{Mason2020}, keeping the door open for its use as a probe of $f_{\mathrm{esc}}^{\mathrm{LyC}}$ at high redshifts.

\subsubsection{Lyman Series}

The Lyman series—spanning from Ly$\beta$ at 1026 \AA, Ly$\gamma$ at 972.5 \AA, Ly$\delta$ at 949.7 \AA, and Ly$\epsilon$ at 937.8 \AA\ down to the Lyman limit at 911.8 \AA—provides another direct probe of neutral hydrogen. Unlike Ly$\alpha$, the higher-order Lyman lines typically lack a strong nebular emission component and instead appear in absorption or as P Cygni profiles, offering a valuable window into the physical state of the ISM and outflows. Their generally lower oscillator strengths at higher energies (e.g., Ly$\beta$–Ly$\epsilon$: 0.079–0.008) lead to lower optical depths, creating more favorable conditions for radiative transfer modeling. The absorption spectra of the Lyman series also present an ideal comparison to metal absorption lines, enabling studies of how metals trace neutral hydrogen. Although the high-energy end of the series often suffers from line blending and foreground contamination from the Milky Way at low redshifts, relatively clean Lyman series absorption features have been observed at redshifts $\sim$0.3 with COS \citep{Henry2015,Carr2023HST}.

\subsubsection{LIS Metal Lines  }

Low-ionization state (LIS) metal lines such as Si II 1190 \AA,1193 \AA, 1265 \AA, 1393 \AA, 1527 \AA\ act as direct probes of the cold clouds ($10^4$ K) that trace the winds of galaxies and ISM.  Si II lines, in particular, have been explored as potential diagnostics of LyC escape \citep{Chisholm2017a}.  Moreover, when combined with the higher ionizations states lines of Si III 1206 \AA\ and Si IV 1393 \AA, 1403 \AA, Si II lines have been used to map the ionization structure of galactic winds and constrain abundance of neutral and total hydrogen \citep{Carr2021a,Xu2022a,Huberty2024}.  Additional lines include N II 1084 \AA, O I 1302 \AA, C II 1334 \AA.  Like the Lyman series, these lines also appear in absorption or as P Cygni profiles.  As metal lines, they typically have much lower column densities than H I, creating ideal conditions for radiative transfer modeling.                                                                                                                                                                                                                                                                                                                                                                                                                                                                                                                                          

Metal lines with fluorescent components, such as Si II 1190 \AA\ (Si II* 1194 \AA) and Si II 1193 \AA\ (Si II* 1197 \AA), offer an ideal opportunity to probe pure absorption profiles, absent blue emission infilling \citep{Prochaska2011,Scarlata2015}.  While resonant emission can fill in absorption wells, fluorescent emission will appear away at longer wavelengths.  In this context, fluorescent lines reveal absorption in the wind and ISM, which includes the redshifted absorption signatures of galactic inflows \citep{Carr2022}.       
                                                                                                                                                                                                                                                                                                                                                                                                                                                                                                                                                                      
\subsubsection{WIS Metal Lines}

Warm-ionization state metal tracers such as N III 990 \AA, C III 977 \AA,  O VI 1032 \AA, 1038 \AA, Si IV 1394 \AA, 1403 \AA, C IV 1548 \AA, 1551 \AA\ offer a potential probe of the warm-hot ($10^{5.5}$ K) phase of galactic winds.  These lines offer a constraint on the most rapidly cooling gas in the CGM, that is prime for re-accretion onto the galaxy, making them ideal probes for how galactic inflows may affect LyC escape \citep{Marques-Chaves2022b}.  When combined with LIS lines, they can be used to further map the ionization structure of the winds, where the majority of mass is typically believed to occupy the cold + warm phases \citep{Kim2020a,Kim2020b,Li2020}.  

Tests against simulations show that the properties of cool galactic outflows–column density ($\sigma \sim 1.3$ dex), opening angle ($\sigma \sim 10$ deg), and flow rates ($\sigma \sim 0.5$ dex)–can be teased out from spectral lines at a resolution $\rm R = 15,000$ and $\rm S/N = 10$.  A spectral resolution of $\rm R = 30,000 - 100,000$ would allow one to resolve outflows down to the turbulence scale of cold clouds \citep{Chen2023}, or the limit at which we can still detect structure.  The restrictions on the warm-hot phase would be less severe, as the higher temperature gases at which these phases probe naturally have higher thermal velocity dispersions which can compete with turbulent line broadening.    

Galactic winds can extend tens to hundreds of kiloparsecs into the CGM, with the cool phase generally reaching shorter distances \citep{Carr2021a,Xu2022a}.  Early observations with COS are limited at nearby redshifts (e.g., \citealt{James2022}) with an aperture of only 1.25 arcseconds allowing one to view a physical scale of just 3.4 kpc at redshift $z = 0.22$.  A 3 arcsecond aperture would allow one to capture the full extent of a 10 kiloparsec wind at redshift 0.22, which is generally enough to capture $> 50\%$ of the Ly halo luminosity \citep{Saldana-Lopez2025}.  

\subsubsection{FUV spectral energy distribution}

Modeling the FUV spectral energy distribution (SED) using BPASS or Starburst99 can provide valuable insights into the relative ages of stellar populations, which in turn can help identify the dominant source of feedback (i.e., radiation vs. supernovae; \citealt{Flury2025,Carr2025LyC}), metallicities, and star formation histories \citep{Chisholm2019}.  Key diagnostic features include the stellar wind lines OVI 1307 \AA\ and N V 1240 \AA, which are particularly sensitive to stellar population age, and the photospheric line C III 117 \AA, which is sensitive to metallicity \citep{Saldana-Lopez2022}.  The aforementioned lines can be well captured at low resolution $R = 1000$ \citep{Saldana-Lopez2022}, but higher resolution $\rm R = 3000-5000$ would be beneficial to capture more lines \citep{Carr2023HST}.  Photospheric lines generally fall between $1000-1800$ \AA\  and wind lines from $1030-1550$ \AA.

\section{Impact on Requirements}

The COS spectrograph currently enables medium-resolution integrated spectroscopy of star-forming galaxies. For example, obtaining medium-resolution spectra of the LyC-leaking starburst J115205 + 340050 ($f_{\mathrm{esc}}^{\mathrm{LyC}} = 13$\%, $z = 0.3419$; \citealt{Flury2022a}) using the COS G130M grating would require approximately 11 hours (12 HST orbits) to reach a signal-to-noise ratio of $\mathrm{S/N} = 10$ at 1000 \AA\, and a spectral resolution of $\sim 75$ km s${}^{-1}$. While feasible, such exposure times are costly, offering an explanation for the scarcity of high-resolution, high-S/N galactic outflow spectra of LyC leakers in the literature.

Using the HWO UV Spectrograph simulator\footnote{\url{https://hwo.stsci.edu/uvspec_etc}}, we estimate that the same $\mathrm{S/N}$ can be achieved—at twice the spectral resolution—in just $\sim$ 48 (27) minutes with a 6 (8)-meter aperture. The FUV channel of the proposed high-resolution spectropolarimeter, POLLUX \citep{Bouret2018}, would be well suited for such observations. With this instrument, high-resolution spectroscopy of galactic winds in LyC-emitting galaxies would become routine. However, to enable truly transformative science, high-resolution spectral imaging is required. A UV IFU would be ideal, providing spatially and spectrally resolved data—an optimal setup for interpretation through radiation transfer modeling (e.g., \citealt{Burchett2021,Erb2023}).  To image SSCs at redshift $z = 0.22$, diffraction limited observations (4.6 (3.5) mas at 1100 \AA\ for a 6 (8)-meter aperture) would be necessary, with a field of view of at least 3 arcseconds.  A multi-object spectrograph could also be valuable, requiring about 20-50 targets per galaxy to capture all LyC emitting clusters.  A summary of the observational requirements is provided in Table~\ref{tab:summary}.

\begin{table*}[!ht]
\caption{Observational Capabilities and Requirements \label{tab:summary}}
\smallskip
\begin{center}
{\small
\begin{tabular}{|p{4cm}|p{1.5cm}|p{1.5cm}|p{4cm}|p{3cm}|}
\hline
\textbf{Capability or Requirement} & \textbf{Necessary} & \textbf{Desired} & \textbf{Justification} & \textbf{Comments} \\
\hline
UV observations (specify wavelength) & Yes & Observed frame: 950--2000~\AA.  & Would allow observations of rest-frame wavelengths $\sim$850--912~\AA\ at $z=0.1$, capturing LyC emission near the Lyman limit. The 912--1500~\AA\ range enables probing of metal lines up to $\sim$1300~\AA\ at $z=0.22$, including Si~II~1190~\AA, 1193~\AA, Ly$\alpha$, etc. 1500-2000 \AA\ would allow us to capture the C~IV~1550~\AA\ doublet, a powerful probe of the warm-hot phase of winds, over redshift range $0 \leq z \leq 0.3$. & ~ \\ 
\hline
Long wavelength observations ($>$1.5 microns) & No & ~ &~& ~ \\
\hline
Timing of observations in different bands & No & & ~ & ~ \\
\hline
High spatial resolution & Yes & Diffraction limited & Required to resolve individual SSCs (1--100~pc), or 0.3--30~mas at $z=0.22$. A diffraction-limited 6(8)-m telescope has 4.6(3.5)~mas resolution at 1100~\AA. Three times this value gives 49(37)~pc at $z=0.22$ and 26(19)~pc at $z=0.1$. & ~ \\
\hline
High spectral resolution & Yes & R = 15,000--100,000 & Minimum of $R=15,000$ to resolve galactic outflow properties such as flow rates, gradients, and column densities.  & $R=30,000$--$100,000$ enables resolving outflows down to the turbulent scale of cold clouds. \\
\hline
Large field of view & Yes & 3 arcsecond aperture & Required to observe a 10 kpc wind at $z=0.22$. & Winds can extend beyond 100~kpc, so this is not an upper limit. 5-10 arcseconds would be exceptional.\\
\hline
Large field of regard & No & & ~ & ~ \\
\hline
Rapid response & No & & ~ & ~ \\
\hline
\end{tabular}
}
\end{center}
\end{table*}



{\bf Acknowledgements.} 
C.~C. is supported by NSFC grant W2433001 and the NSFC
Talent-Introduction Program.  R.~C. acknowledges in part financial support from the start-up funding of Zhejiang University and Zhejiang provincial top level research support program.
\bibliography{author.bib}

\end{document}